\documentclass{article}
\usepackage{spconf,amsmath,graphicx}
\usepackage{multirow}
\usepackage{kotex}
\usepackage{adjustbox}
\usepackage{xcolor}
\usepackage{array}
\usepackage{footnote}
\usepackage{makecell}
\usepackage{svg}

\title{Room adaptive conditioning method for sound event classification\\in reverberant environments}
\name{Jaejun Lee$^{1}$, Donmoon Lee$^{1,3}$, Hyeong-Seok Choi$^{1}$, and Kyogu Lee$^{1,2}$}
\address{$^{1}$Music and Audio Research Group, Department of Intelligence and Information,
\\$^{2}$Artificial Intelligence Institute, Seoul National University, $^{3}$Cochlear.ai\\}

\begin{document}
\ninept
\maketitle

\begin{abstract}
Ensuring performance robustness for a variety of situations that can occur in real-world environments is one of the challenging tasks in sound event classification. One of the unpredictable and detrimental factors in performance, especially in indoor environments, is reverberation. 
To alleviate this problem, we propose a conditioning method that provides room impulse response (RIR) information to help the network become less sensitive to environmental information and focus on classifying the desired sound.
Experimental results show that the proposed method successfully reduced performance degradation caused by the reverberation of the room.
In particular, our proposed method works even with similar RIR that can be inferred from the room type rather than the exact one, which has the advantage of potentially being used in real-world applications.
\end{abstract}

\section{Introduction}
\label{sec:intro}

Sound event classification (SEC) is a task that automatically categorizes audio clips into labels that match their acoustic content.
The SEC system is being studied on a variety of topics with the aim of application in real-world,
such as abnormal event detection \cite{valenzise2007scream, foggia2015reliable}, urban sound recognition \cite{park2014sensing}, or domestic sound recognition \cite{krstulovic2018audio}.
However, the application of the SEC system is difficult due to the various acoustic distortions that occur when applied to a real-world environment. For example, background noise other than the desired event obscures important sounds, and various sounds overlap due to reverberation effects.

Unfortunately, the effort to apply the SEC system to a real-world environment is still focused on securing noise robustness against noise \cite{foggia2015audio, mcloughlin2015robust, ozer2018noise}. One of the major reasons for this is the complexity of the reverberation properties. In particular, a simple mathematical operation can produce a sound with a specific reverberation applied using its characteristics, but the opposite is difficult without knowing the same reverberation information. To overcome this, in the field of speech recognition, studies have suggested methods that directly model complex reverberations through neural networks. The representative examples are dereverberation methods that remove reverberation characteristics from speech signals \cite{kinoshita2017neural, wang2018investigating}. However, in the case of SEC, it is difficult to directly apply the dereverberation method because the target event sound is ambiguous and more diverse than the voice.

Meanwhile, recent studies have suggested various ways of using the encoded information in neural networks. It starts with simply using the high-level information of the pre-trained network \cite{johnson2016perceptual, dumoulin2016learned}, and is also used as a way to give the desired direction for network training while simultaneously learning various information from multiple inputs \cite{dumoulin2018feature, perez2018film, kim2017dynamic}. In particular, applying this method as a feature level has the advantage of being able to intentionally train the network by blocking and passing specific information through the gate.


In this study, we propose a novel method that suppresses the effect of reverb motivated by recent network conditioning methods. Our proposed method simultaneously takes the audio input used in a typical SEC and an additional reverberation characteristic, the room impulse response (RIR), of the target room environment. It differs from previous studies is that these additional inputs were not used to focus on specific information, but to lessen concentration on specific information.
The proposed method has the advantage of potentially being used in real-world applications since RIR is relatively easy to acquire and for a given SEC service, the location rarely changes.

Our contributions are followings: 
First, we suggest a test environment to evaluate the performance degradation caused by reverberation. We then present a network training method that suppresses specific environmental factors. Our proposed method can be modularized so that it can be additionally trained from the existing network, and that it works even by using only similar RIRs that can be estimated in the environment as revealed in the experiment.

\section{Background}

\subsection{Room Impulse Response}
Under a linear time-invariant condition, room impulse response (RIR) completely describes the acoustic path including source sound propagation and room reflections \cite{kuttruff2016room}.
It means sound distorted by reverberation can be modeled by the convolution operation of exact RIR and source audio.
Even in the same room, RIR varies depending on the location of the source audio and the interference of objects. It makes non-availability of deconvolution which is inverse transformation of convolution, because the result of deconvolution without exact RIR is significantly distorted from source audio.

Reverberation time ($T_{60}$) and direct-to-reverberant ratio (DRR) are well-known representative measures that show characteristics of the RIR. $T_{60}$ is defined as the time it takes for the source level to decay 60 dB after the source has ceased. The DRR is the energy ratio between sound come from the direct path of source audio and other reflected paths of it. As the distance to the source increases, reverberant energy remains constant, but the direct path energy decays by 6 dB per doubling the source distance, results in DRR is inversely proportional to source distance \cite{richards1980reverberations}. Note that source audio is distorted more as $T_{60}$ increases and DRR decreases.

\subsection{Feature-wise Transformation}
In machine learning, conditioning means context-based processing that processes source information in the context of another.
Feature-wise transformation is one of the conditioning methods which is simply done by element-wise affine transformations between source information and auxiliary information \cite{dumoulin2018feature}.
Also, such the transformation with non-linear activation operates like a gating mechanism that allows the conditioning information to select which features are passed forward and which are not. In the previous study \cite{perez2018film}, the network that trained with the image (source information) and the text (auxiliary information) tended to focus more on the object related to the text and zeroed out the rest of the image.
Despite the simplicity of its structure, the feature-wise transformation has proven its effectiveness in previous studies using various modalities such as image, audio, and text \cite{perez2018film, dumoulin2016learned, kim2017dynamic}.


\begin{table}[!t]
\renewcommand{\arraystretch}{1}
\centering
\begin{adjustbox}{width=0.47\textwidth}
\tiny
\begin{tabular}{|c|c|c|c|c|c|}
\hline
Test set &  \begin{tabular}[c]{@{}c@{}}RIR\\dataset\end{tabular} & \begin{tabular}[c]{@{}c@{}}Room type\end{tabular} & \begin{tabular}[c]{@{}c@{}}\# of\\RIR\end{tabular} & \begin{tabular}[c]{@{}c@{}}$T_{60}$(s)\end{tabular} & Notation \\
\Xhline{4\arrayrulewidth}
\multirow{7}{*}{\begin{tabular}[c]{@{}c@{}}Simulated\\ Test set\end{tabular}} &
AIR & Booth & 12 & 0.27 & \textit{R027} \\
& WDR & CR7 & 360 & 0.29 & \textit{R029} \\
& AIR & Office & 12 & 0.39 & \textit{R039} \\
& MARDY & - & 73 & 0.55 & \textit{R055} \\
& AIR & Lecture & 24 & 0.68 & \textit{R068} \\
& AIR & Stairway & 78 & 0.77 & \textit{R077} \\
& QMUL & Classroom & 130 & 1.34 & \textit{R134} \\ \hline
\multirow{2}{*}{\begin{tabular}[c]{@{}c@{}}Recorded\\ Test set\end{tabular}}
& - & Corridor & 1 & 0.20 & \textit{Record1} \\
& - & Boardroom & 1 & 0.22 & \textit{Record2} \\ \hline
\end{tabular}
\end{adjustbox}
\caption{The specification of the simulated and recorded test sets. The test sets are renamed with the notation using its $T_{60}$.}\label{tab:testset}
\end{table}

\section{Proposed method}

The proposed method is a conditioning network using the room impulse response (RIR) of the target room. The source audio which is distorted by the room reverberation is transformed by the RIR of that room.
For the conditioning method, feature-wise transformation including scaling and biasing operation is used, because it is assumed that if it is possible to make the network focus on the information of input related to another input, the opposite would also be possible. This means that it will be possible to make the network focus on the information that is not related to another input which is the reverberation characteristic of the target room. The RIR is given so that the network concentrates more on sound event characteristics only, ignoring obstruction of room reverberation.
Concretely, the proposed conditioning network consists of three embedding blocks. One for source audio, and two for the RIR of scaling and biasing operation each.
With the input source audio $x$ and the RIR of that room $r$, the transformed embedding output $y$ is defined as,
\begin{equation}
y = \sigma(H(x)*R_s(r))+R_b(r)
  \label{eq1}
\end{equation}
where $H$, $R_s$, and $R_b$ are the embedding block of $x$, the embedding block for scaling and it for biasing, respectively. Also, $*$ and $\sigma$ are element-wise multiplication and non-linear operation such as \textit{ReLU}, respectively. Then $y$ is provided for classification network after another non-linear activation.
The proposed method has the advantage of being a separable module that can be attached to the conventional deep learning-based SEC models.

\begin{figure}[!t]
  \centerline{\includegraphics[width=0.9\columnwidth]{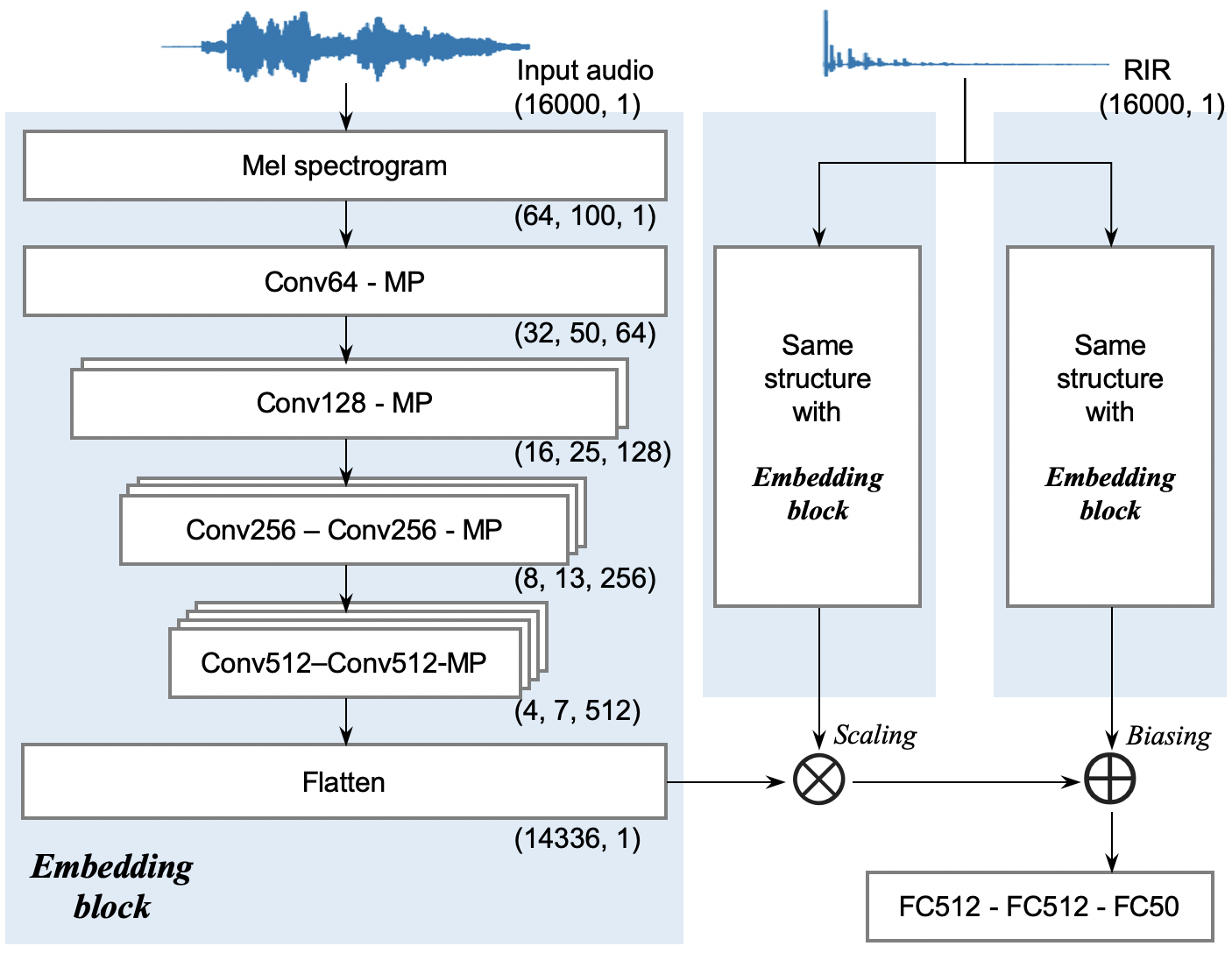}}
\caption{The network architecture. \textit{Conv}, \textit{MP} and \textit{FC} mean convolution, max-pooling and fully connected layer each with the filter dimension.
The numbers in parentheses below each layer represent its output dimension.
}
\label{fig:architecture}
\end{figure}

\section{Experiment}

\subsection{Dataset}
Our experiments focus on how a classification model trained in a clean environment works in real-world reverberant rooms. In this context, we used \textit{Real World Computing
Partnership} (RWCP), a dataset containing various kinds of sound events recorded in an anechoic chamber. The test set consists of a clean room, simulated reverberation rooms, and recorded reverberation rooms.

\textbf{Clean test set} The clean test set is test set of the entire data from RWCP, 50 classes with 80 clips were used in this experiment following previous studies \cite{mcloughlin2015robust, ozer2018noise, dennis2012image}. The clean test set consists of 20 clips assigned to each class, and the rest are used for training.

\textbf{Simulated test set} Simulated test sets are made by convolving real-world impulse response (IR) into the test data of SEC dataset. As a real-world IR, Aachen impulse response (AIR) \cite{jeub2009binaural}, multichannel acoustic reverberation database at York (MARDY) \cite{wen2006evaluation}, Queen Mary University of London (QMUL) \cite{stewart2010database}, and Westdeutscher Rundfunk (WDR) \cite{stade2012spatial} are used. A brief summarization for each simulated test set is provided in Table \ref{tab:testset}.

\textbf{Recorded test set} The clean test set is re-recorded by playing in a fixed location in two separate environments. The first is a two-walled \textit{corridor} and the second is a four-walled \textit{boardroom}. Brief reverberation information for each room is summarized in Table \ref{tab:testset}. The sound source audio was played from the speaker rather than the real sound source, but the reverberation was generated in the real-world environment, so we assumed it was closer to the real-world environment compared to the simulated test sets.

\begin{figure*}[!t]
  \centerline{\includegraphics[width=0.92\textwidth]{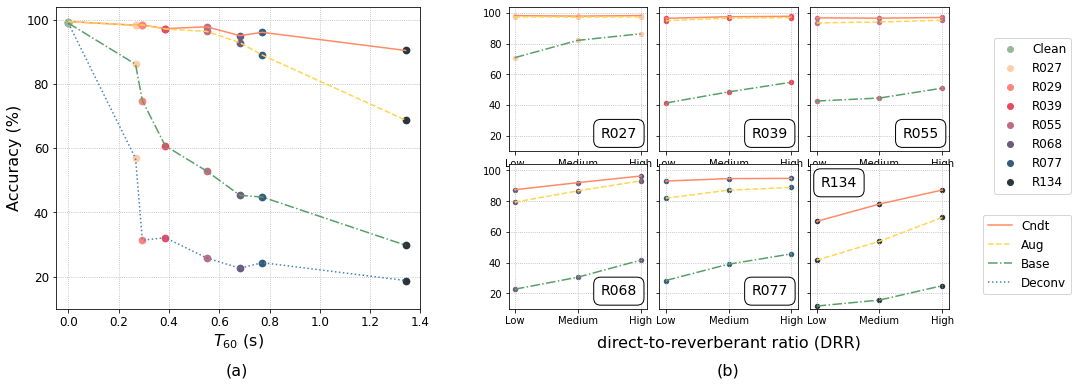}}
\caption{The results of each model in the original clean test set (\textit{Clean}) and the simulated test sets. (a) shows performance related to the $T_{60}$ and (b) shows performance related to the DRR in the chosen six rooms.}
\label{fig:result1}
\end{figure*}

\subsection{Network architecture}
The network architecture is based on the \textit{VGGish} network \cite{hershey2017cnn}, which is known to be effective for extracting audio features. We used four convolution blocks structure of \textit{VGGish} for the embedding block of our network.
Since all the inputs have the same modality which is audio, the same structure is used for the embedding block of the input source audio and the two embedding blocks of the RIR for scaling and biasing, but the parameters are not shared.
Details of the embedding block structure are given in Figure \ref{fig:architecture}.
All convolution layers have a filter size (3x3) and (1,1) strides with padding for the same output size. Batch normalization and \textit{ReLU} activation were applied after every convolution layer. All max-pooling layers have a pooling size (2x2) and (2,2) strides with padding for the same output size. Scaling and biasing layers with \textit{ReLU} activation are provided for conditioning, and followed by a two fully connected (FC) layers with the 512 units. It is then connected to the output layer.

The raw audio input is converted to mel spectrogram with 32 ms window length, 10 ms hop length, 125 Hz the lowest frequency, and 7,500 Hz the highest frequency. The input of the network has the shape of (64, 100) corresponding to the number of frequency bands and frames.

\subsection{Implementation details}
The dataset has an audio length of about 1 second with short silence before and after the sound. We only used the first 1 second of audio and zero-padded if it is shorter than 1 second.
All audio was resampled to 16 KHz and subjected to maximum normalization.

The performance of the proposed method was evaluated as an average of 10 trials and the validation set was randomly selected with 20\% of the training set in each trial.
For training, was used a batch size of 64 and the maximum number of epochs were set to 100. Adam \cite{kingma2014adam} optimizer with a learning rate of 0.0001 were used, and we chose the model which has the lowest loss in the validation set. 
The experiment is implemented in 
\textit{Tensorflow} \cite{tensorflow2015-whitepaper}.

\subsection{Training strategies} \label{subsec:training strategies}
In the experiments, we compare four different training strategies which are \textit{Base}, \textit{Deconv}, \textit{Aug}, and \textit{Cndt}. The \textit{Base} is trained using only the original train set with 3,000 data.

The \textit{Deconv} has the same training strategy as \textit{Base}, but at the inference time, it has the RIR deconvolution operation at the front of the network. Since the exact RIR cannot be obtained in the real-world target environments, at the inference time, we randomly selected the RIR at different points within the same room rather than the exact RIR.

The \textit{Aug} is trained using an augmented train set that convolved random virtual RIR with the original train set for every epoch. The reason for using virtual RIR is that the real-world RIR is not diverse enough to apply the augmentation method, on the other hand, the virtual RIR has the advantage that can be created indefinitely with various reverberation characteristics.
The virtual RIR is generated using the image method \cite{allen1979image} and its implementation\footnote{https://github.com/ehabets/RIR-Generator}.
We set up nine virtual rooms, each with a different size and different reverberation time between 0.15 and 0.7 seconds to cover various room conditions. We generated 100 RIRs at random points in each room, resulting in a total of 900 virtual RIRs.

\textit{Cndt} is trained using an augmented train set and applied the proposed conditioning method. The exact virtual RIR that convolved with the input audio is given as a RIR input pair. For the RIR pair of the original clean train data, an unit impulse signal (discrete delta function) that does not change the signal after convolution is provided. Also, similar to \textit{Deconv}, at the inference time, we randomly selected the RIR at different points within the same room rather than the exact RIR.

\section{Result and discussion}
\subsection{Performance degradation in the simulated test sets} \label{subsec:result1}
Figure \ref{fig:result1}(a) shows performance on the four different training strategies on the original clean test set (\textit{Clean}) and the simulated test sets (\textit{R027}$\sim$\textit{R134}).
\textit{Base} shows over 98\% accuracy on the \textit{Clean} which has about 0 seconds $T_{60}$. However, performance degrades in simulated test sets that are distorted by reverberation. For the rooms that have over 0.6 seconds $T_{60}$, accuracy is degraded under 50\%. It is observed that reverberation affects SEC performance significantly and the degradation is intensified as the $T_{60}$ increases.

\subsection{Non-availability of the deconvolution algorithm} \label{subsec:result2}
\textit{Deconv} in Figure \ref{fig:result1}(a) shows a significance performance drop as $T_{60}$ increases.
As mentioned in Section \ref{subsec:training strategies}, we used the RIR of a different point in the same room, which is similar to the exact one, but not precisely the same. The results show that performance is worse than the \textit{Base} model and it implies that in the real-world, even if we can get a RIR of the target room, the deconvolution algorithm is non-available.

\begin{figure}[!t]
  \centerline{\includegraphics[width=0.9\columnwidth]{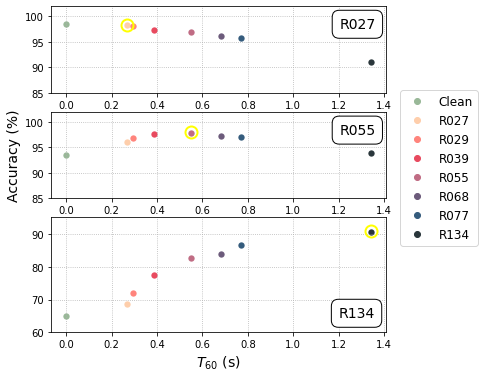}}
\caption{The results of the fake conditioning experiment with \textit{Cndt} in three rooms \textit{R027}, \textit{R055}, and \textit{R134}. Each dot point represents the performance conditioned with each fake room for the fake conditioning, and the dot point with a yellow circle represents the performance of the true conditioning.}
\label{fig:result2}
\end{figure}

\subsection{Proposed conditioning method} \label{subsec:result3}
\textit{Aug} in Figure \ref{fig:result1}(a) shows over 98\% accuracy on the test sets with short $T_{60}$ (\textit{Clean}, \textit{R027}, \textit{R029}). However, as the $T_{60}$ increases, the performance decreases. Especially test sets that have $T_{60}$ over 0.7 second, which is the maximum $T_{60}$ value of the augmented train set, show an intensified performance drop. The result indicates that it is difficult to achieve stable performance only using the augmentation technique, especially when the $T_{60}$ of the target room is long.

Comparing with \textit{Aug}, \textit{Cndt} shows statistically significant performance improvement in four test sets with long $T_{60}$ (\textit{R055}, \textit{R068}, \textit{R077}, \textit{R134}) ($p$ $<$ 0.01). It suggests that the proposed conditioning method complements the performance of the \textit{Aug}, and the longer the reverberation time of the test set, the greater the effect.
The results of the \textit{R055}, \textit{R068} indicate that the proposed conditioning method gives an additional enhancing effect even if the reverberation time of the target rooms is considered in the train set by augmentation technique. Also, the results of the \textit{R077}, \textit{R134} indicate that with the unknown target rooms in the training phase, the proposed method can improve performance using non-training information of target rooms. It means that it does not need to re-train the classifier, so it can reduce the amount of data required for training by alleviating the need to force the model to have the generalized performance to all rooms.
For scaling and biasing operation for conditioning, similar to \cite{perez2018film}, using only the scaling layer showed more performance gain than using only the biasing layer, but using both layers achieved better results. An extensive analysis of what information each operation is conditioning is left as future work.

\subsection{Performance related to the DRR} \label{subsec:result4}
Figure \ref{fig:result1}(b) shows the performance of the chosen six rooms related to the direct-to-reverberant ratio (DRR) of each room. We divided the RIRs of each room into three groups (\textit{Low}, \textit{Medium}, \textit{High}) related to the DRR value and evaluated performance using the simulated test set per room and per DRR group.

\textit{Base} in Figure \ref{fig:result1}(b) shows that performance increases as DRR increases and it implies that if the distance to source audio increases, the performance will decreases. Similar to the results of Figure \ref{fig:result1}(a), comparing with \textit{Aug}, \textit{Cndt} shows statistically significant performance improvement in all three DRR groups in four test sets with long $T_{60}$ (\textit{R055}, \textit{R068}, \textit{R077}, \textit{R134}) ($p$ $<$ 0.01). It implies that the proposed conditioning method can improve performance not only in various $T_{60}$ room environments but also in various DRR points.

\subsection{Fake conditioning experiment} \label{subsec:result5}
We conducted the fake conditioning experiment that conditions using the RIR of rooms different from the room of the source audio, which means a pair of inputs is mismatched intendedly.
Figure \ref{fig:result2} shows its results on three test sets that are chosen in consideration with a variety of $T_{60}$ lengths. According to the results of \textit{R027}, the performance degradation is minimal when conditioning with RIR of rooms having $T_{60}$ similar to \textit{R027}, which is relatively short, but the performance decreases as the $T_{60}$ increases. On the other hand, the results of \textit{R134} show a relatively slight performance degradation when conditioned with the RIR in rooms with long $T_{60}$ similar to \textit{R134}, but with shorter $T_{60}$, the degradation is intensified. In the case of \textit{R055}, the degradation is minimal when conditioning with the RIR of rooms having an intermediate $T_{60}$ similar to \textit{R055}, but is intensified when conditioning with RIR of rooms having a relatively short or long $T_{60}$. The results of the fake conditioning experiments show that the proposed conditioning network tends to improve performance with RIR that has a similar reverberation time of the target room, and it suggests that with only the reverberation time information, it can be used to improve the SEC performance in that room, even if the exact RIRs of the target room are unknown.

\subsection{Evaluation on recorded test set} \label{subsec:experiment5}
Table \ref{tab:result3} shows the results on the recorded test sets which is closer to real-world reverberant environments. The RIR for each test set was simply acquired by recording impulse sound at a random point in each room.
We found that the results show the same tendency as the simulated test sets. \textit{Base} shows a significant performance drop in each real-world reverberant room, and \textit{Aug} mitigates the degradation to some extent. The results of the proposed \textit{Cndt} shows significant performance enhancement compared to other models. It suggests that our proposed method works not only in the simulated environments but also in the real-world reverberant environments.

\begin{table}[!t]
\centering
\footnotesize
\begin{tabular}{c|cccccccccccccc}\hline
\begin{tabular}[c]{@{}c@{}}Model\end{tabular} & 
\begin{tabular}[c]{@{}c@{}}Clean\end{tabular} &
\begin{tabular}[c]{@{}c@{}}Record 1\end{tabular} &
\begin{tabular}[c]{@{}c@{}}Record 2\end{tabular} \\ \Xhline{4\arrayrulewidth}
\begin{tabular}[c]{@{}c@{}}Base\end{tabular} & 98.92 & 56.17 & 58.67\\
\begin{tabular}[c]{@{}c@{}}Aug\end{tabular} & 99.42 & 78.95 & 72.99\\
\begin{tabular}[c]{@{}c@{}}Cndt\end{tabular} & 99.43 & 84.38 & 78.47\\ \hline
\end{tabular}
\caption{Classification accuracy (\%) on the recorded test sets.}
\label{tab:result3}
\end{table}

\section{Conclusion} \label{sec:conclusion}
In this study, we proposed a performance enhancement method for the SEC system applicable to real-world room environments. Evaluations using test data modeled the real-world reverberant rooms confirmed that the performance degradation worsened with increasing the reverberation time and decreasing the direct-to-reverberant ratio. To address this issue, we proposed a conditioning network using the room impulse response (RIR) of the target room. Experimental results show that the proposed method mitigates the performance degradation under various reverberation characteristic conditions. Moreover, the proposed conditioning method tends to improve performance even with the approximate RIR that have a similar reverberation time of the target room. It means that even if the exact RIR of the target room is unknown, our proposed method is still beneficial.
Since the effectiveness of the method of adaptively utilizing the reverberation information of the target environment has been verified, a real-time calibration of the SEC system using more various adaptive information such as noise is left as future work.

\vfill\pagebreak



\bibliographystyle{IEEEbib}
\bibliography{strings,refs}

\end{document}